\documentclass[aps,nofootinbib,superscriptaddress,11pt]{revtex4-1}
\usepackage{amssymb}
\usepackage{amsmath}
\usepackage{graphicx}
\usepackage{color}
\usepackage{slashed}

\newcommand{\req}[1]{Eq.\,\eqref{#1}}

\newcommand\beq{\begin{equation}}
\newcommand\eeq{\end{equation}}
\newcommand\bea{\begin{align}}
\newcommand\eea{\end{align}}

\begin{document}

\title{Finding quantum effects in strong classical potentials}
\author{B. Manuel Hegelich}
\author{Lance Labun}
\affiliation{Department of Physics, University of Texas, Austin, Texas 78712 U.S.A.}
\author{Ou Z. Labun}
\affiliation{Department of Physics, University of Texas, Austin, Texas 78712 U.S.A.}
\affiliation{Department of Physics, University of Arizona, Tucson, Arizona 85721 U.S.A.}
\date{14 July 2016}

\begin{abstract}
The long-standing challenge to describing charged particle dynamics in strong classical electromagnetic fields is how to incorporate classical radiation, classical radiation reaction and quantized photon emission into a consistent unified framework.  The current, semiclassical methods to describe dynamics of quantum particles in strong classical fields also provide the theoretical framework for fundamental questions in gravity and hadron-hadron collisions, including Hawking radiation, cosmological particle production and thermalization of particles created in heavy-ion collisions.  However, as we show, these methods break down for highly relativistic particles propagating in strong fields. They must therefore be improved and adapted for the description of laser-plasma experiments that typically involve the acceleration of electrons. Theory developed from quantum electrodynamics, together with dedicated experimental efforts, offer the best-controllable context to establish a robust, experimentally-validated foundation for the fundamental theory of quantum effects in strong classical potentials.
\end{abstract}
\maketitle

\section{Introduction: classical and quantum radiation}

Although the photoelectric effect showed that radiation is fundamentally a quantum process, long-wavelength ($\lambda \gg \lambda_e =\hbar/m_ec$ where $m_e$ is the electron mass) radiation is equally well described by solving Maxwell's equations with a classical current as source.  This classical approximation works because, in weak fields, the energy of the probe $\hbar\omega =h c/\lambda\ll m_ec^2$ is insufficient to excite the quantum structure of the electron.  The electron Compton wavelength thus usually provides a natural length scale separating photons that should be treated as quantum ($\lambda\lesssim \lambda_e$) and photons that can be treated as approximately classical ($\lambda\gg\lambda_e$).  However, by creating sufficiently strong fields, we can probe quantum dynamics with potential energy, rather than kinetic energy, with qualitatively different consequences.

The most remarkable and earliest-recognized consequence is Klein's 1929 discovery that when the potential difference exceeds the mass gap, $e\Delta V> 2m_e$ where $e$ is the elementary charge, the transmission coefficient for scattering is greater than unity \cite{Klein:1929zz}.  The potential emits particles at a rate that depends on the details of the potential but is always nonzero when $e\Delta V>2m_e$ \footnote{Klein obtained his result using the Dirac equation, but before antiparticles were understood or positrons discovered.}.  This spontaneous particle creation was the first example of a nonperturbative effect in quantum theory, because it required solving the interaction with the potential to all orders, rather than expanding in a power series in the coupling $e$.  Klein studied a step potential $V(z)\sim \Delta V\,\Theta(z)$, in which case the current, corresponding to pair creation, is proportional to the height of the potential step.  In more realistic models, the potential depends on the spacetime coordinates more smoothly, varying over a length scale greater than the electron Compton wavelength, $\lambda\gg \lambda_e$ and the rate of pair production is exponentially suppressed, proportional to $\exp(-\frac{\pi E_c}{|\vec E|})$ where 
\beq\label{Ecdefn}
E_c=\frac{m_e^2c^3}{e\hbar}=1.32\times 10^{18}~\mathrm{V/m},
\eeq
is known as the critical field and $|\vec E|$ is the local value of the electric field.  Paradoxically, long-wavelength classical fields were found to generate short-wavelength $\sim\lambda_e$ radiation.  

Klein's ``paradox'' is the prototype to help understand phenomena and fundamental questions in domains other than electrodynamics.  In quantum chromodynamics, the theory describing quark and gluon interactions inside hadrons, it is thought to help explain the high multiplicity and thermalization of particles created in the early stages of heavy ion collisions \cite{Andersson:1983ia,Kharzeev:2006zm,Gelis:2014qga,Gelis:2015kya}.  In gravity, the same physics underlies Hawking radiation \cite{Brout:1995rd,Kim:2007ep} and particle creation in the expanding universe \cite{Anderson:2013zia,Anderson:2013ila}, both predictions subject to ongoing investigation and debate.  Magnetic fields of the corresponding strength,
\beq
B_c=\frac{1}{c}E_c=4.41\times 10^9\,\mathrm{T},
\eeq
 are believed to exist around neutron stars \cite{Harding:2006qn}.   Even with such broad impact and long interest, spontaneous particle production has yet to be experimentally verified.

Strong fields break the classical approximation in another more subtle way.
Already Lorentz noticed that conservation of momentum requires that the charged particle recoils from the radiation it emits according to Maxwell's equations, but that the recoil is not accounted for in the Lorentz force \cite{ALD}.  This ``radiation reaction'' problem can be perturbatively corrected by subtracting the lost momentum from the Lorentz force \cite{ALD,LandauLifshitz}, but even this solution is not complete:  First, the perturbative approach breaks down when the momentum in the classical radiation is comparable to the momentum in the radiating electron. The onset of nonperturbative classical radiation approximately coincides with the condition that the field strengths in the electron rest frame are $\simeq E_c,B_c$ \cite{Hadad:2010mt}, and in this same regime, high-energy quantized radiation becomes equally important \cite{Ritus:1985}. Second, the momentum absorbed by the electron is not removed from the accelerating field. Wheeler and Feynman solved the second problem by revising the Green's function to be the average of the retarded and advanced functions thus communicating the momentum absorbed by the electron back to the source of the field \cite{Feynman:1965jda}.  

Feynman's propagator is a basic prescription of quantum field theory, and means radiation reaction is always consistently incorporated in quantum theory, but is not compatible with classical radiation theory.  In principle, instead of using the classical limit, one could simulate the full quantum dynamics as is done in lattice studies of quantum chromodynamics.  In practice, such simulations are impossible for plasma systems of interest: to resolve the quantum dynamics, the simulation's grid spacing should be at least the order the electron Compton wavelength $\sim 10^{-12}$\,m and the simulation volume should be large enough to incorporate tens to hundreds plasma wavelengths, which is $\lambda_{pl}\simeq 10^{-6}$\,m, thus involving a spacetime mesh of $\gtrsim (10^7)^4$ points.  This wide separation of length scales is what ensures the classical limit remains an important tool to describe plasma physics even as quantum effects become important.

The challenge to describing particle dynamics and radiation in strong fields is thus to build a consistent and systematic framework to treat classical (long-wavelength) and quantized (short-wavelength) radiation where they are both important.
The best controllable theoretical and experimental opportunities to explore these issues are in quantum electrodynamics (QED), facilitated by new ultra high power lasers \cite{Dunne:2008kc,Hegelich:2014tda}.
While spontaneous particle emission, requiring fields close to $10^{18}$ V/m, is too rare an event to be measured in current laser facilities, we now study electron dynamics in moderately strong ($10^{14}$--$10^{15}$ V/m) fields to see the effect of radiation reaction and perturbative and nonperturbative quantum radiation processes that occur in the presence of the strong classical field \cite{DiPiazza:2011tq}.

The current theory for these processes is founded upon the study of particle creation.  Here we first review the methods in their original context of spontaneous particle creation and then critically examine their applicability to anticipated laser experiments.  In this well-studied framework for QED in strong fields, average particle creation rates and light-by-light scattering involving low-energy photons ($\hbar \omega\ll m_ec^2$) are consistently and accurately described because the long-wavelength fields do not resolve the short-wavelength quantum fluctuations.  For electrons and high-energy photons with $\hbar\omega\gtrsim m_ec^2$, scattering events that involve energies of order the electron mass require a quantum description, because roughly speaking they probe the structure of the electron.  For the field strengths in anticipated experiments on the Texas Petawatt ($I\simeq 3\times 10^{22}$ W/cm$^2$, or $|\vec E|\gtrsim 10^{15}$ V/m)  and at ELI ($\gtrsim 10^{23}$ W/cm$^2$, or $|\vec E|\gtrsim 5\times 10^{15}$ V/m) , such events are relatively rare, and one expects that classical dynamics suffice to determine evolution between events.

Our goal is the accurate and reliable prediction of high-intensity laser experiments, and we discuss here how to construct a consistent and systematic theory for the plasma dynamics, starting with single-particle radiation phenomena.  As such, we consider the experimental range of single-particle conditions, rather than attempting to optimize the signal from the current, untested calculational approaches.

With this goal in mind, any description of the interactions between quantized particles requires a systematic investigation of the perturbation theory.  Based on well-established features of quantum field theory especially involving massless gauge fields such as photons, we argue that QED in strong classical fields (laser 4-potentials $eA^{\mu}\gg m_e$) gives rise to $\mathcal{O}(1)$ corrections to probabilities of events.  We describe the impact of these corrections on predictions involving real high-energy electrons and photons and the contrast from low-energy vacuum phenomena (i.e. no quantized particles present) or non-relativistic electrodynamics.  With its broader applications, a framework to describe the interactions of particles with strong, classical fields is an unsolved problem that would have fundamental impact both within and beyond plasma physics.  We introduce the concepts needed to improve on the current theory and how we expect the development of more complete and systematic theory to impact the broader questions in related fields arising from Klein's ``paradox.''

\section{Spontaneous pair production and semiclassical methods}

Although Klein's calculation relied on a simple scalar step potential, it captured the essential physics.  Shortly afterward, Sauter showed that particle emission occurs in a constant electric field, which with specific choice of gauge corresponds to a linear potential $A^\mu=(A_0,\vec 0)$ with $A_0=-|\vec E|z$ \cite{Sauter:1931zz}.\footnote{We use natural units, $\hbar=c=1$, restoring them explicitly where adding insight.}  For a constant field of infinite extent, the potential difference is $2m_e$ over a length $\Delta z=2m_e/|e\vec E|$.  Using intuition from the quantum mechanical barrier scattering problem, the mass gap $2m_e$ between positive and negative frequency states implies a classically forbidden region with width $\Delta z$ in which the wavefunction of the electron decays exponentially.  The decay length is proportional to the energy of the state and so is  maximized for an electron at rest $\sim 1/m_e$.  Tunneling through the barrier is therefore exponentially suppressed $\sim e^{-m\Delta z}=e^{-\frac{2m_e^2}{|e\vec E|}}$.\footnote{One may get the exponent correct by massaging the argument further, but in general, beyond the relevant dimensionless ratios of parameters, such qualitative arguments will not reproduce order 1 prefactors.}  
The exponent exhibits the critical field \req{Ecdefn}
which sets the scale for pair production to become an order 1 effect and therefore also estimates the breakdown scale of the theory.

To obtain their results, Klein and Sauter solve the Dirac equation (or Klein-Gordon equation for a scalar) in the presence of a c-number electromagnetic potential, $A_{\rm cl}^\mu$,
\beq\label{DiraceqnwithAcl}
(i\slashed{\partial}-e\slashed{A}_{\rm cl}(x)-m)\psi(x)=0.
\eeq
The subscript `cl' stands for classical because the potential $A^\mu(x)$ is nondynamical, and incorporating it into the solution of the Dirac equation assumes that the potential varies slowly relative to the dynamics of the quantized electron, i.e. the following condition is satisfied:
\beq\label{semiclassiccond}
\frac{\partial V(x)}{V(x)}\ll \frac{\partial\psi(x)}{\psi(x)} 
\qquad \Leftrightarrow \qquad
k_{\rm cl}^\mu \ll p_e^\mu.
\eeq
Expanding each field in Fourier modes shows it is equivalent to the wavelength of the ``classical'' modes being much larger than the wavelength of the quantized modes.
In this case, a single photon of the classical field does not resolve the quantum structure of the electron.  This condition is consistent with separately solving the classical electromagnetic field dynamics.  As discussed below, it allows the equations of motion for the vacuum expectation value of the current (arising from spontaneously produced partices) and slowly-varying electromagnetic fields to be closed and solved self-consistently at the long-wavelength scale.  The approximation Eqs.\,\eqref{DiraceqnwithAcl},\eqref{semiclassiccond} also provides a starting point to calculate rates of quantum processes compatible with classical laser-plasma simulations, but we shall discuss processes with electrons (and high-energy photons) in later sections.

Equation \eqref{semiclassiccond} is a necessary condition; it does not determine useful applications.  The approximation Eqs.\,\eqref{DiraceqnwithAcl},\eqref{semiclassiccond} is applied in cases that the potential is leading order.   In contrast, standard perturbation theory assumes the particle is free at leading order and interactions are suppressed by a small coupling.  In QED, that means solving the free Dirac equation $(i\slashed{\partial}-m)\psi=0$ to obtain a basis of plane wave solutions and adding interactions with photons as a perturbative expansion in $\alpha=e^2/4\pi\hbar c$.  However, if $eA^\mu \gtrsim p^\mu_e$,  perturbation theory in $eA^\mu$ breaks down.  Instead, the classical potential should be taken into account to all orders, summed into a new leading order by organizing perturbation theory as an expansion around solutions to \req{DiraceqnwithAcl}.  
The solutions of \req{DiraceqnwithAcl} often display non-polynomial dependence on $eA_{\rm cl}$, though they can be expanded in powers of $eA_{\rm cl}$.  For a plane-wave potential (discussed more extensively in the next section), the first terms in the expansions are explicitly verified to agree with a perturbative calculation of the electron scattering from the potential \cite{Lavelle:2013wx}.  

Treating the electromagnetic field as classical, via Eqs.\,\eqref{DiraceqnwithAcl},\eqref{semiclassiccond}, is a necessary and consistent approximation scheme when the electron field is integrated out and only dynamics in vacuum are considered.  The process of integrating out is equivalent to and often implemented by using the leading order equation of motion for the relevant degrees of freedom to remove them from the (quantum) generating functional.  Thus, solving \req{DiraceqnwithAcl} allows one to remove electrons as an explicit degree of freedom and determine the effect of electron fluctuations on the dynamics of long-wavelength electromagnetic fields.

However, as $|\vec E|=E_c$, the potential difference is $m_e$ over the Compton wavelength of the electron $\lambda_e=1/m_e$, and the relevant timescale for pair creation approaches $\hbar/\alpha m_e c^2=1.8\times 10^{-19}$ s \cite{Labun:2008re}, indicating that the field-particle dynamics are no longer adiabatic.  Indeed, the rapid conversion of field energy to particles has been suggested to limit the electric field strength achievable \cite{Fedotov:2010ja,Bulanov:2010gb}. We will explain this breakdown more thoroughly below.  Other than these emergent dynamics, treating the electromagnetic field as classical provides a good leading-order description of the pair creation dynamics performs well, due largely to the smallness of the electromagnetic coupling $\alpha=\frac{e^2}{4\pi \hbar c}\simeq \frac{1}{137}$.

This procedure of integrating out the electrons was first achieved by Heisenberg and Euler \cite{Heisenberg:1935qt}, who unified Sauter's investigation with the first calculations of light-by-light scattering \cite{Euler:1935zz}, by calculating the correction to the energy of a constant electromagnetic field due to quantum fluctuations of electron-positron pairs.  In so doing, they obtained the first effective potential and first low-energy effective field theory, a framework of great importance that we will explain below.  The Heisenberg-Euler effective potential shows that long-wavelength $\lambda\gg\lambda_e$ light-light interactions can be expanded in a power series with succeeding terms suppressed by the ratio $|\vec E|/E_c$:
\beq\label{EHVeff}
V_{\rm eff}\simeq\frac{\alpha}{\pi}|\vec E|^2\sum_{n=1}^{\infty}a_n\left(\frac{|\vec E|}{E_c}\right)^{2n}
\eeq
(suitably generalized to be parity conserving and depend only on the field tensor $F^{\mu\nu}$ and its dual in the presence of magnetic fields).
Clearly the power series fails to converge as $|\vec E|\to E_c$.  This breakdown signals the onset of nonperturbative physics \cite{Dunne:1999uy,Labun:2008qq}, which in this case we already know is the probability of particle creation being near unity.

Schwinger advanced the calculational technology by obtaining the effective potential \req{EHVeff} in a gauge-invariant approach that also allowed a straightforward definition of its imaginary part \cite{Schwinger:1951nm}
\beq\label{ImVeff}
\mathrm{Im}\,V_{\rm eff} = \frac{\alpha}{2\pi^2}|\vec E|^2\sum_{n=1}^\infty \frac{1}{n^2}e^{-n\pi E_c/|\vec E|}
\eeq
$V_{\rm eff}$ is complex due to the particle creation instability; to create the electric field we must do work to separate charges to $z\to \pm\infty$ and the field can be screened and reduced by creating electron-positron pairs.  The imaginary part limits the radius of convergence of the power series \req{EHVeff} and ensures that the series is only asymptotic.  

Moreover, recalling this energy balance points to a small inconsistency in the framework: the electric field in the calculation is taken as prescribed, and the energy required to create the pair is not removed from the field.  This can be effectively corrected by incorporating the pair creation dynamics into a local mean-field current $\langle j^\mu(x)\rangle$ that sources the electromagnetic field via Maxwell's equation
\beq\label{semiclassicalmaxwell}
\langle j^{\mu}(x)\rangle=\partial_\nu F^{\nu\mu}(x)\,,
\eeq
The vacuum expectation value of the current $\langle j^\mu\rangle$ is calculated under the approximation Eqs.\,\eqref{DiraceqnwithAcl},\eqref{semiclassiccond} and evolved forward in time using \req{DiraceqnwithAcl} according the Schwinger-Keldysh formalism for non-equilibrium quantum dynamics \cite{Cooper:1989kf,Kluger:1992,Kluger:1998bm}.
The resulting system of kinetic equations is closed and consistent at the long-wavelength scale $\lambda\gg\lambda_e$.  The classical field and mean particle number dynamics can also be followed with greater temporal resolution using a real-time statistical formulation \cite{Hebenstreit:2013qxa}.

This system of equations is consistent as long as the timescale for a quantum event (pair production) is much smaller than the timescale for changes in the classical field, which is estimated by the plasma frequency
\beq\label{semiclassicaldynamicscondition}
\tau_q \ll \tau_{pl}=\frac{2\pi}{\omega_{pl}}\,.
\eeq
Numerical study of spontaneous pair production dynamics indicate that $\tau_q\sim 1/E_{p}$, where $E_{p}$ is the energy of the created particle \cite{Kluger:1998bm,Anderson:2013zia}.  To estimate the plasma timescale, we consider near-critical fields creating pairs at a rate given by the first term in the series \req{ImVeff} \cite{Cohen:2008wz,Labun:2008re}
\beq\label{pairrate}
\frac{dN_{pairs}}{d^4x}=\frac{\alpha}{8\pi^3}|\vec E|^2 e^{-\pi E_c/|\vec E|} 
\eeq
Then using the classical definition of the plasma frequency for pairs created by the field, $\omega_{pl}^2=e^2 n_e/m_e$ with the density of charges $n_e=2n_{pairs}=2\tau_{pl}\frac{dN_{pairs}}{d^4x}$, we obtain 
\beq
\tau_{pl} = \frac{1}{m_e}\left(\frac{2\pi^2}{\alpha}\frac{m_e^2}{|e\vec E|}\right)^{2/3}e^{\frac{\pi E_c}{3|\vec E|}}
\eeq
Although crude, solving this estimate shows that \req{semiclassicaldynamicscondition} is satisfied by orders of magnitude even up to $|\vec E|=E_c$, due to the smallness of $\alpha$.  At this level, the closed system of pair creation and the electron-positron plasma dynamics backreacting on the field appears to be consistently classical as desired.

A more insidious inconsistency takes over in the limit $|\vec E|\to E_c$ however.  The conversion rate of field energy into rest mass $d\langle u_m\rangle/dt$ is characterized by a kinetic timescale, the ``materialization time''  \cite{Labun:2008re} 
\beq\label{taumat}
\frac{1}{\tau_{mat}}=\frac{2}{|\vec E|^2}\frac{d\langle u_m\rangle}{dt}\simeq\frac{1}{\tau_{e}}|\vec E|^2e^{-\pi E_c/|\vec E|}
\qquad
\tau_{e} = \frac{\pi^2}{\alpha m_e}\simeq 1.7\times 10^{-18}~\mathrm{s}
\eeq
Already at $|\vec E|\sim 0.4E_c$, the materialization time is less than 1 femtosecond, indicating that a significant fraction ($\sim 1/e$) of the laser energy would be converted into pairs in less than one period of an optical laser field (see also \cite{Fedotov:2010ja,Bulanov:2010gb}).

This implies $\sim 1/e$ of the total system's energy is in the pair plasma, and therefore the plasma-induced electric and magnetic fields are similar in magnitude to the laser field and varying on the time and length scale of the plasma density 
\beq\label{pairplasmalengthscale}
\Delta t,\Delta L\sim \left(\frac{dN_{pairs}}{d^3x}\right)^{-1/3}\sim\left(\tau_{mat}\frac{dN_{pairs}}{d^4x}\right)^{-1/3}\sim 9/m_e,
\eeq
an estimate agreeing with the real-time simulations under the same approximation \cite{Hebenstreit:2013qxa}.  Equation \eqref{pairplasmalengthscale} shows that the mean distance between particles is much smaller than the classical electron radius $9/m_e\ll 1/\alpha m_e\simeq 137/m_e$, and \req{taumat} thus estimates the field scale at which pair creation and field dynamics are strongly coupled to each other.  Subsequent pair creation events are significantly affected by the presence of nearby charges, which indicates that the vacuum (no particles present) and slowly-varying (\req{semiclassiccond}) conditions in the estimates of pair creation are violated.  An additional condition for this method to describe the particle and field evolution in laser experiments is therefore that the materialization time is greater than one laser period, $\tau_{mat}>T_{\rm laser}$.

To establish that the framework of \cite{Cooper:1989kf,Kluger:1992,Kluger:1998bm,Hebenstreit:2013qxa} is a good approximation scheme to describe spontaneous pair production, it only remains to check that quantum corrections are under control, i.e. the expansion in $\alpha$.  The two-loop ($\alpha^2$) contribution to pair production has been calculated explicitly, showing that it only becomes important at $E\sim 60E_c$ \cite{Ritus:1977,Reuter:1996zm,Fliegner:1997ra}.  In sum, the  framework is valid and can be self-consistently closed in moderately strong, but smaller than critical fields $|\vec E|\lesssim 0.4E_c$.  On the other hand, recall also that classical radiation emission also becomes nonperturbative (a leading order correction to electron dynamics) at this field strength \cite{Hadad:2010mt}.  Thus, this method breaks down near the critical field, and a new theory is needed to describe dynamics at or above this field scale.

The same framework describes cosmological particle creation and backreaction on black holes due to Hawking radiation. In these cases, the classical potential is gravitational and the quantum dynamics are similarly obtained by solving for the wavefunctions in the presence of the potential.  Backreaction is included by calculating the source current (there the energy-momentum tensor) in the same approximation scheme (Eqs.\,\eqref{DiraceqnwithAcl},\eqref{semiclassiccond}) and plugging it back into the field equations, as in \req{semiclassicalmaxwell}.  

Recent progress has also been made on the statistics of particles produced, toward improving the understanding of thermalization and multiplicity of particles created in heavy-ion collisions.  Here, the early stages of a collision between two heavy ions, usually Au and Pb nuclei, are modeled by considering that hard collisions between partons inside the nucleons are relatively rare but respond to the long-wavelength gluon fields sourced by partons in the other nucleus.  Solving the classical equations of motion (analogs of Maxwell's equations for non-abelian theory) for these long-wavelength gluon fields suggests the creation of extended color fields in the aftermath of the collision \cite{Andersson:1983ia,Kharzeev:2006zm,Gelis:2015kya} that can decay via spontaneous emission. 

\section{High-intensity laser experiments}

So far, we have studied how a specific nonperturbative process, spontaneous pair production, is described in quantum field theory by approximating the strong electromagnetic field as classical.  In this study, the initial state of the system contains no particles, and the subsequent evolution is only defined in the mean field approximation, which means considering dynamics at the long-wavelength scale of the classical field and averaging over the charge density at shorter distance.  We need a different theoretical framework to predict the dynamics of \emph{particles} interacting with a strong classical field and the relevant experimental observables such as (final state) energies and radiation emitted by particles strong laser fields.  Indeed, the same considerations apply to the single particle dynamics at the onset of pair creation, when only a few particles are present and are accelerated by the strong field to radiate and possibly go on to create more pairs, as suggested by \cite{Fedotov:2010ja,Bulanov:2010gb}. To develop a useful theory however, we first discuss the conditions and measurable results of experiments.

Most experiments involve a high intensity ($I>10^{21}$ W/cm$^2$, $|\vec E|>10^{13}$ V/m) laser pulse striking a solid or gas target at rest in the laboratory frame.  Current laser systems achieve peak intensities of $10^{22}$ W/cm$^2$, and planned facilities will reach $10^{23}-10^{24}$ W/cm$^2$, still much smaller than the intensity corresponding to the critical field \req{Ecdefn}
\beq\label{Icdefn}
I_c = \epsilon_0 cE_c^2 = 1.29 \times 10^{29}~\mathrm{W/cm}^2\,.
\eeq
To maximize the field strength seen by electrons, the density of electrons is chosen to be above the critical density
\beq\label{ncrdefn}
n_e > n_{cr} = \frac{m_e\omega_{laser}^2}{e^2}
\eeq
obtained by setting the plasma frequency equal to the laser frequency.  Setting $n_e>n_{cr}$ ensures the laser frequency is below the low-frequency cutoff for electromagnetic waves that propagate in the plasma, and consequently a large fraction of laser energy is absorbed or reflected, creating electric and magnetic fields in the plasma of the same order of magnitude as the incident laser field.  (Note that $n_{cr}$ is unrelated to the critical field strength, being purely classical with no $\hbar$.)

The experiments are equipped with detectors that typically measure the particle momentum in a small region of phase space.  For example, a high-energy particle spectrometer may cover a significant range in one momentum component (e.g. along the beam axis) but have a small angular acceptance, with particles only less than a few milliradians from the beam axis entering the detector.  On the other hand, detectors with larger angular acceptance are typically designed for the low-energy ($\hbar\omega\ll m_ec^2$) classical radiation.  This situation is understandable given the fore-going focus on particle acceleration schemes.  

To study quantum radiation and particle production processes, we will need to adapt diagnostics in everyday use at high energy particle colliders such as Jefferson Lab or KEK, not to mention larger facilities such as the Relativistic Heavy Ion Collider (RHIC) and the Large Hadron Collider (LHC).  These detectors are suited to the measurement of high energy particles in larger volumes of phase space and the extraction of rare signal events amid a much larger number of background particles, capabilities also important for seeking quantum effects in high-intensity laser experiments.  However, their designs are less prepared for the high flux of radiation from a typical laser-plasma experiment.  For example, for Texas Petawatt parameters, current models predict $\gtrsim 10^{10}$ photons with energies $\gtrsim 1$ MeV entering 20-degree by 20-degree solid angle around the beam axis and distributed in time over an interval less than 1 picosecond.  This flux is orders of magnitude greater than tracked by the LHC detectors, which now handle $\gtrsim 100$s of particle tracks every 100 picoseconds (the crossing rate of bunches at the collision points).  While particle density would be mitigated by larger distance to the detector (as will probably be necessary anticipating larger detector infrastructure), dedicated development of new detector systems will be a necessary component of future high-intensity laser experiments, especially those seeking quantum effects.

One experiment so far has provided proof-of-principle that high-energy particle detectors are useful to diagnose laser-particle interactions.  The SLAC E-144 experiment successfully collided the linear accelerator's 46.6 GeV electron beam with a moderate intensity laser ($I\simeq 5\times 10^{17}$ W/cm$^2$).  They detected nonlinear Compton scattering $N\gamma+e\to \gamma+e$ and pair production, achieving agreement with predictions in the weakly-nonlinear regime $N<10$, where the strong classical field is not yet dominant \cite{Bula:1996st,Bamber:1999zt}.

\section{Processes and observables for electrons in strong laser fields}

Predictions are derived by two methods: 1) analytic calculations of single electron dynamics, utilizing the above-described approximation for ; and 2) numerical particle-in-cell (PIC) simulations, implementing a reduced model of quantum emission processes and/or radiation-reaction modified equations of motion.  The two differ in several important aspects.  

Analytic theory calculations are equipped to investigate quantum interferences in the amplitudes, which in principle could be engineered to enhance the rate for a specific process to occur.  For example, a sequence of $N$ (identical) electric field pulses is predicted to increase the probability of spontaneous pair creation by $N^2$, analogous to $N$-slit diffraction \cite{Akkermans:2011yn}.  However, such calculations require that the electromagnetic field is known at all points in spacetime, and thus they typically exclude any dynamical plasma effects that affect the profile of the input laser, classical radiation and backreaction of the calculated processes.

PIC simulations achieve the complement: developed initially to predict the nonlinear laser-plasma dynamics, they include the backreaction on the classical laser field due to the collection motion and radiation by the plasma.  Due to the computational expense however, PIC simulations typically use one simulation (quasi-)particle to represent $N\gg 1$ physical electrons or ions.  They can be adapted to special-case single-particle studies and take into account evolving particle multiplicities, such as in \cite{Elkina:2010up}.  However in general, PIC codes are not suited to incorporate the stochastic character of quantum processes and instead implement simplified versions.  A common but important simplification is that photons are emitted only exactly collinear to the electron 3-momentum, so as to exclude stochastic momentum dispersion within a simulation quasi-particle.  Recent work has suggested this simplification is reasonably accurate, provided one can average over a large number $\sim 10^3$ simulations \cite{Harvey:2014qla}; however statistical uncertainties in the simulations (and experiments) remain unknown. Moreover, because the long-wavelength electromagnetic fields are solved on a grid, PIC simulations intrinsically introduce a new length scale, the grid cell size, that separates classical electromagnetic dynamics from any quantized photon emission.  This separation is artificial and should be removed systematically using techniques described below.

Even so, one can learn about the dynamics by well-designed calculations with the laser field treated as prescribed, and we summarize a few relevant results here.  Most recent results cited here rely on solutions to the Dirac equation \req{DiraceqnwithAcl} with a plane-wave type potential in the transverse gauge: $A^{\mu}_{\rm cl} = A^{\mu}_{\perp}(t-z)$, the wave propagates in the $+\hat z$ direction, hence a function only of $t-z$, and its polarization is in the transverse ($x,y$)-plane.  The solution is known as the Volkov solution \cite{Wolkow:1935zz,Brown:1964zzb} and is invariant only under a restricted class of gauge transformations (see \cite{DiPiazza:2013vra} for a generalization of the Volkov solution).  The common procedure is then calculate QED diagrams with the classical field incorporated into the basis of electron states used to form matrix elements and electron correlation functions.  Real processes have only been calculated at tree-level, without loop corrections.  The one-loop (order $\alpha$) self-energy and photon polarization tensor have been obtained \cite{Ritus:1985}; however, their divergent parts contribute renormalization and thus are visible only in higher order corrections, which have not been studied, and the real processes represented by their imaginary parts are equivalently obtained from tree-level diagrams.

Toward the construction of a more systematic quantum theory, it is useful to show that this method reproduces properties of QED perturbation theory.  One of the more important results is that low-momentum radiative corrections factorize and exponentiate  \cite{Dinu:2012tj,Ilderton:2012qe}.  The low-energy, low-intensity limit of the photon emission probabilities agree with the classical limit \cite{Ritus:1985}.  Additionally, the Ward-Takahashi identity, ensuring gauge invariance of amplitudes involving electron loops constructed from Volkov states, holds \cite{Meuren:2013oya}, and the optical theorem explicitly checked for the one-loop polarization function for photons \cite{Meuren:2014uia}.  These results, including the analysis of the propagator \cite{Lavelle:2013wx}, arise from the fact that Volkov solution is a Wilson line: a gauge-covariant, path-ordered line integral that can be expanded into an infinite sum of couplings to the here-classical gauge potential.  To see this, note that the Volkov solution for an electron can be written
\beq
\psi_V(x) = \sum_p W_p e^{-ipx}u_p
\eeq
where $u_p$ satisfies the free Dirac equation $(\slashed{p}-m)u_p=0$ and the prefactor satsifies \cite{Ritus:1972ky}
\beq
(i\slashed{\partial}-e\slashed{A}_{\rm cl})W_p\hat{\mathcal{O}}=W_pi\slashed{\partial}\hat{\mathcal{O}}
\eeq
for any operator $\hat{\mathcal{O}}$.  Thus, somewhat more formally, $W_p$ satisfies an operator equation $(i\slashed{\partial}-e\slashed{A}_{\rm cl})W_p=0$, which is the equation defining a Wilson line for the potential $A_{\rm cl}^\mu$.

Beyond these field theoretic results, a small tome's worth of articles have been published predicting phenomenological signatures of high-energy photon emission and pair production, which we shall not review comprehensively here.  In a monochromatic plane-wave field, momentum conservation requires that an electron can only radiate photons in integer multiples of the plane-wave wave vector, $k^\mu_{\rm out}=Nk^\mu_{\rm laser}$.  $N=1$ corresponds to perturbative Compton scattering, and $N>1$ is often called non-linear Compton scattering because it requires absorbing $N$ quanta from the classical field and becomes more probable as the laser $a_0\gg 1$.  Many studies focus on the kinematical consequences of short (few cycle) laser pulses, which broadens the momentum distribution of photons for the electron absorb \cite{Heinzl:2009nd,DiPiazza:2010mv,Seipt:2010ya,Mackenroth:2010jr,Seipt:2012tn,Harvey:2012ie,Mackenroth:2012rb}.
Pair conversion by photons and pair emission by electrons propagating in the field has been studied under similar conditions
\cite{Meuren:2014uia,Heinzl:2010vg,Ilderton:2010wr,Titov:2012rd,King:2013osa,Nousch:2012xe,Meuren:2014kla,Jansen:2015idl}.  Even the effective neutrino-photon coupling and the axial anomaly have received attention \cite{Shaisultanov:1997bc,Shaisultanov:2000mg,Gies:2000tc,Meuren:2015iha}.

One phenomenologically important fact is that for $a_0\gg 1$ pair creation in a general plane-wave field is well-approximated by convolving the local constant crossed-field probability of creation with the classical dynamics of the electrons \cite{Meuren:2015mra}.  This outcome may be understoond by noting that $a_0$ is the inverse of the Keldysh parameter \cite{Keldysh:1965,Schutzhold:2008pz}, which  when $1/a_0\sim 1$ roughly indicates that the frequency of the classical field is important.  Conversely when $1/a_0\ll 1$, the probability approaches the constant field result.  Thus, for $a_0\gg 1$ laser fields, varying only in one lightcone direction, the process is local: there are no nonlinear vacuum polarization effects in a single plane-wave of arbitrary spectral composition \cite{Schwinger:1951nm} (a second wave-vector must be introduced to break the symmetry), and pair creation in a light-like electric field is a local event \cite{Ilderton:2015qda}.  This property is particularly important for constructing a systematic quantum theory; if it were not true, we would have to worry about the length scale over which non-local correlations in the field could impact short-wavelength (quantum) dynamics.

\section{Effective field theory for strong field processes}

With the goal of constructing a predictive theory, what is the salient difference between pair creation in vacuum and electron dynamics in strong fields?  In the study of spontaneous pair creation or light-by-light scattering with no real particles present, there are two length scales: the wavelength of variation of the classical field and the Compton wavelength of the electron.  The condition \req{semiclassiccond} is the statement that these length scales are widely separated.  Physically, the electron is a heavy particle, whose fluctuations are point-like relative to the variation of the (classical) field.  The electron can therefore be ``integrated out'', that is the degrees of freedom can be removed from the theory as fluctuating too quickly to be resolved by the long wavelength dynamics.  

Integrating out the electron field to one-loop order yields the Heisenberg-Euler effective potential for long wavelength ($\lambda\gg\lambda_e$) electromagnetic fields, which is thus an expansion in both the ratio $\omega/m_e$, controlling the importance of derivative corrections, and the QED coupling $\alpha$, controlling the importance of additional loop corrections (see \cite{Ritus:1977} for the two-loop action).  Although a complex-analytic expression is known to all orders in the fields, applying the effective potential to any real process for electromagnetic fields in vacuum utilizes the expansion in powers of $1/m_e^4$ (the fourth power due to the fact that only field invariants $\sim F^2$ can appear), shown in \req{EHVeff}.  As described above, the expansion in powers of $1/m_e^4\sim 1/E_c^2$ breaks down at the critical field coinciding with the onset of significant pair creation -- which anyway breaks the no-particles-present condition.  \footnote{The first two terms in the power series $n=2,3$ are a very good approximation up to $|\vec E|\simeq E_c$, and the divergence of the power series manifestly coincides with the imaginary part becoming order 1 \cite{Labun:2008qq}.  Pure magnetic fields are stable, suggesting one might apply the Heisenberg-Euler effective potential to nonlinear dynamics of constant magnetic fields with magnitude greater than $B_c$.  However the dynamics obviously introduces time-derivatives on the fields, and the likely application (magnetized compact object atmospheres) involves photons and plasmas with particles in the keV energy range.}

In contrast, processes involving real high-energy $\hbar\omega\gtrsim m_ec^2$ photons or electrons in the classical field involve at least one additional length scale, the de Broglie wavelength of the particle.  It is simpler from now on to refer to momentum scales $p,k\sim \hbar c/L$.  For photons with momentum $k\ll m_e$, electrons are again heavy and can be integrated out.  The Euler-Heisenberg effective potential thus suffices to calculate probabilities for vacuum birefringence, photon splitting and four-wave mixing.  Photons with momentum $k\sim m_e$ and low-momentum electrons however resolve fluctuations around the scale $m_e$ and a full quantum theory with dynamical electron and photon degrees of freedom is necessary. 

The case of greatest interest phenomenologically is photons and electrons with momentum $k,p\gg m_e$, as arise in the final state of a typical acceleration experiment.  In this case, there is a large hierachy of momentum scales
\beq\label{laserhierarchy}
E+p_z \gg |eA_{\rm cl}^\mu| \gg m_e \gg \omega_{laser}
\eeq
where $eA_{\rm cl}$ is the amplitude of the classical laser potential.  The use of $E+p_z=p_+$ the lightcone momentum conjugate to the lightcone coordinate $x_-=t-z$ is natural due to the planewave symmetry of the background.  Current high intensity laser systems on which the next experiments will be conducted have
\beq
a_0=\frac{|eA_{\rm cl}^\mu|}{m_e} \sim 10-350
\eeq
at their peak.
\footnote{The vector potential $A^{\mu}$ is well-defined for a high intenity field, because both the field intensity $|\vec E|^2$ and Fourier decomposition (distribution in $k$-space) can be measured with precision smaller than the absolute values.  Consequently,  the canonical degrees of freedom $A^{\mu}(x)$, $\partial_t A^{\mu}(x)$ commute to very good approximation, the field is classical, the photon occupation number is not fixed.}

\begin{figure}
\centerline{\includegraphics[width=0.3\textwidth]{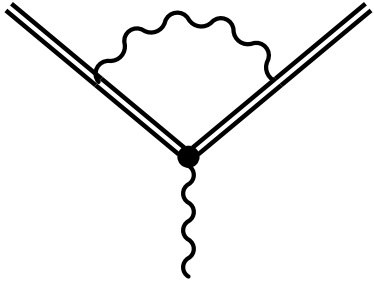}}
\caption{One loop correction to the electron-photon vertex.  The double line indicates the dressed propagator, which incorporates the classical potential to all orders.  Details of the calculation will be presented elsewhere.\label{fig:1loopvertex}}
\end{figure}

This hierachical separation of momentum or length scales is important to account for in quantum processes, because perturbation theory generically introduces corrections that are parametrically large, involving logarithms of ratios of the physical scales.  Although the coupling constant is small for QED, a widely-separated hierachy, such as $E+p_z\sim 10^3m_e\sim 10^8\omega_{laser}$, enhances quantum corrections to becoming relevant.  As an example, we consider the radiative corrections to the rate of photon emission.  The largest difference between the tree-level and 1-loop rates is a double logarithm involving the scales seen in \req{laserhierarchy}\cite{Labun:2016}
\beq\label{sudakov}
\frac{\Gamma^{\rm (1-loop)}(e\to e\gamma)}{\Gamma^{\rm (tree)}(e\to e\gamma)}\underset{-q^2\gg m_e^2}{\simeq} 1-\frac{\alpha}{4\pi}\ln\frac{-q^2}{m_e^2}\ln\frac{-q^2}{E_d^2}
\eeq
where $-q^2=-(p-p')^2$ is the squared 4-momentum change by the electron.  The coefficient is obtained by evaluating the one-loop correction to the electron-photon vertex shown in Fig. \ref{fig:1loopvertex}, with the laser field included to all orders by using the dressed propagator for the electron.  

One can guess the form of \req{sudakov} realizing that, even in the presence of a strong classical field, the short-distance behaviour of the amplitude must reproduce zero-field ($A_{\rm cl}^\mu\to 0$) QED.  Fortunately, amplitudes for QED in classical laser fields do have the same short-distance behaviour as in zero-field QED.\footnote{These features were not made manifest in the computation by \cite{Morozov:1981pw}, which focussed on the asymptotics for $\chi\gg 1$, and will elaborated by us in a dedicated publication.}  The classical field affects only the long-distance dependence, here encoded in $E_d$.

The momentum scale $E_d$ is the energy resolution of a detector and is an infrared cutoff distinguishing radiation from the electron \cite{Peskin:1995ev}; in other words it measures the precision to which the electron momentum can be known.  Here $E_d$ is set by the classical radiation, which having $\omega_{cl.rad.}\ll m_e$ is a continuous process relative to photon emission.  We therefore estimate this scale as the cyclotron frequency of the electron in the laser field $\omega_{cl.rad.}\sim \omega_{cyc}=|e\vec B|/p$.  This can be written Lorentz invariantly by going to the instantaneous rest frame of the electron, where the field strength is $|\vec B'|$ and
\beq
E_d\sim\omega_{cyc}=\frac{|e\vec B'|}{m_e} = \chi
\eeq
with
\beq\label{chidefn}
\chi^2 = \frac{1}{m_e^4}p_\mu eF^{\mu\nu}eF_{\nu\lambda}p^{\lambda} = p\cdot P_{laser}
\eeq
the Lorentz invariant presenting the center of mass energy in the collision between the electron and the classical field (the momentum density of the field is multiplied by the Compton volume of the electron $\lambda_e^3=1/m_e^3$ to form a momentum).  The dimensionless $\chi/m_e$ controls the magnitude of quantum effects, as shown by explicit calculation, and the limit $\chi/m_e\to 1$ implies the electric field seen by the electron in its rest frame is equal to the critical field \req{Ecdefn}. 

\begin{figure}
\centerline{\includegraphics[width=0.3\textwidth]{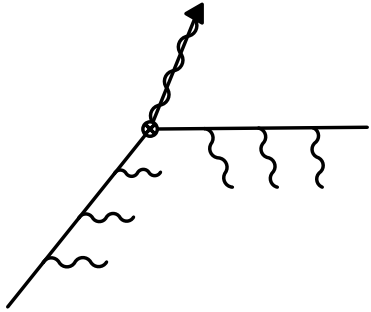}}
\caption{Real radiative corrections that correspond to the large logarithm \req{sudakov}, that is the emission of (any number of) lower-energy photons between the laser field frequency and the energy of the detected high-energy photon.\label{fig:legcorrections}}
\end{figure}

Equation \eqref{sudakov} represents perturbative corrections to the emission probability that are present both in zero-field QED and QED with a classical laser potential.  The role of the laser field is to make single-photon emission, the tree-level amplitude, possible.  The physical reason for the size of these corrections is the widely-separated hierarchy of scales \req{laserhierarchy} and will arise under any consideration of high-energy electrons recoiling with $-q^2\gtrsim m_e^2$.  Intuitively, one can think of the associated emission (or absorption) of a large number of lower-energy photons, as shown in Fig. \ref{fig:legcorrections}.  The energy of these lower-energy photons can lie anywhere between the low-energy region $E_d\sim \chi$ and the electron's recoil, which encompasses a large region of phase space when $-q^2\ggg\chi^2$.  These lower-energy photons are distinguishable from classical field, which in most applications, using the quasi-constant approximations, consists of only the zero-mode; indeed photons at $E_d$ momentum scale coincide with the classical radiation predicted by the Lorentz force.  

PIC simulations necessarily introduce another momentum scale, the inverse of the grid spacing, which sets an upper cutoff on the frequency of classical electromagnetic fields solved on the grid.  For accurate single electron trajectories, this momentum scale should be chosen high enough to resolve the cyclotron frequency of electrons in the laser field \cite{Gonoskov:2014mda,Arefiev:2016zqq}, and thus it naturally separates the classical and quantum radiation processes.  The quantum emission rates should be evaluated with explicit knowledge of this scale, which is only possible with the methods we have briefly introduced in this article.

This does not require the fractional change in the electron momentum to be large: for a 50-MeV electron co-propagating with $a_0\sim 10^2$ laser, $\chi/m_e\sim 10^{-6}$, emitting a 5 MeV photon is subject to a $\sim 6-15\%$ correction.  The emission probability for a photon with energy $k^\mu\sim 50$ MeV (still only one percent the electron's classical kinematic momentum $\sim a_0^2m_e$) is corrected by nearly 50\%. Since photons emitted in the energy range $k^\mu\sim 500$ keV $-50$ MeV are a primary observable of interest in laser experiments, and simulations with the current PIC models suggest they are produced in observable numbers, it is necessary to ensure predictions of the photon emission spectrum and angular distribution are correct in this region.  Corrections of the magnitude suggested \req{sudakov} are more than enough to perturb the photons from their expected locations in the experimental apparatus and miss the (typically small-aperture) detectors completely.  We are concerned with being able to verify predictions of QED effects: a detector merely ``seeing something'' does nothing to validate the theory.

For an electron with $\gtrsim 50$ MeV energy co-propagating with the laser, the field in its rest frame is reduced by the Lorentz factor, and $\chi\sim 10^{-5}$.  Then \req{sudakov} implies that the emission probability for a photon with similar energy $k^\mu\sim 50$ MeV is corrected by nearly 50\%.  For a moderate energy photon $k^\mu\sim m_e$, the first logarithm in \req{sudakov} is replaced by a number of order 1, and the correction is 2-5\%, depending on the electron momentum.  Note that this is equal or larger than corrections implied by classical models of radiation reaction-corrected dynamics \cite{Hadad:2010mt}.

\section{Conclusions}

Finding quantum effects in strong fields will require substantial combined theoretical and experimental efforts.  Here, we have only touched briefly on the supporting computational infrastructure, that is necessary to connect the theory, describing high-energy, short-distance quantum dynamics, to the experiments, involving low-energy, long-distance classical dynamics.  However, it is clear that more complete, systematic method to relate quantum and classical dynamics, must be incorporated into the numerical simulations that are used to interpret laser-plasma experiments.

We have provided a preview of results proving that additional theory, beyond the tree-level amplitudes with the laser field treated as non-dynamical, is required to accurately describe quantum radiation processes by electrons in strong fields.  Moreover, these same calculational methods underlie several outstanding issues in theoretical physics, from particle production in heavy-ion collisions to Hawking radiation in black holes, highlighting the necessity of a more complete theory of quantum and classical radiation and the transition between.  As physics relies on experimental verification, it is of pressing importance to develop the theory in parallel with on-going high-intensity laser experiments, so as to provide predictions of directly relevant experimental observables.  To this end, we anticipate that the experience offered by high energy hadron collisions will be invaluable, both in designing and building detectors and in defining observables to help test the theory.

Concerted, coordinated effort in this program will not only address some of the longest-standing problems in plasma physics (radiation reaction and spontaneous pair production) but also provide the first experimental foundation for fundamental theoretical issues in gravity and strong nuclear interactions.

\end{document}